\documentclass[3p,12pt]{elsarticle}

\usepackage{lineno,hyperref,amsmath}
\usepackage{amsthm}
\usepackage{amssymb}
\usepackage{color}
\usepackage{graphicx}
\modulolinenumbers[5]

\newcommand{\fref}[1]{Figure \ref{#1}}
\newcommand{\sref}[1]{\S \ref{#1}}

\newcounter{mnote}
\begin{document}

\begin{frontmatter}

\title{Metastable transitions in inertial Langevin systems: what can be different from the overdamped case?}

\author[GaTMath]{Andre N. Souza}
\ead{andre.souza@gatech.edu}
\author[GaTMath]{Molei Tao}
\ead{mtao@gatech.edu}
\address[GaTMath]{School of Mathematics, Georgia Institute of Technology, Atlanta, GA 30332 USA}

\begin{abstract}
 Metastable transitions in Langevin dynamics can exhibit rich behaviors that are markedly different from its overdamped limit. In addition to local alterations of the transition path geometry, more fundamental global changes may exist. For instance, when the dissipation is weak, heteroclinic connections that exist in the overdamped limit do not necessarily have a counterpart in the Langevin system, potentially leading to different transition rates. Furthermore, when the friction coefficient depends on the velocity, the overdamped limit no longer exists, but it is still possible to efficiently find instantons. The approach we employed for these discoveries was based on (i) a simple rewriting of the Freidlin-Wentzell action in terms of time-reversed dynamics, and (ii) an adaptation of the string method, which was originally designed for gradient systems, to this specific non-gradient system.

\end{abstract}


\end{frontmatter}


\section{Introduction}
The Langevin equation
\begin{align}
\label{lang}
\begin{array}{r@{}l}
 d X &= V dt  \\
 M d V &= -\Gamma  V  dt + f(X) dt+  \epsilon  \Gamma^{1/2} d W
 \end{array}
\end{align}
can exhibit richer behaviors than its overdamped limit
\begin{align}
\label{over}
 d X &= \Gamma^{-1} f(X)dt+  \epsilon  \Gamma^{-1/2} d W,
\end{align}
where $M \in \mathbb{R}^{n \times n}$ is a positive definite mass matrix, $\Gamma \in \mathbb{R}^{n\times n}$ is a positive definite friction coefficient matrix, the variables $X, V \in \mathbb{R}^n$, $\epsilon \in \mathbb{R}$, and $W$ is an $n$-dimensional Wiener process. When $\epsilon = 0$ the equations are deterministic and play a special role in the $\epsilon \rightarrow 0$ limit. We also refer to the $\epsilon = 0$ case by the same names, but when ambiguity arises we will add the term ``noiseless" to distinguish the two cases.

Let $\| \cdot \|_{A}$ denote a weighted norm,
\begin{align}
\| \mathbf{x} \|_A = \sqrt{\mathbf{x}^T A \mathbf{x}},
\end{align}
where  $\mathbf{x} \in \mathbb{R}^n$ and $A \in \mathbb{R}^{n \times n}$ is a positive definite matrix.
 Freidlin-Wentzell theory shows the transition from one point in the state space to another of \ref{lang} or \ref{over} is characterized by the minimizer of the following actions 
\begin{align}
\label{lang_func}
L[\mathbf{x} ] &= \int_0^T \| M \mathbf{\ddot{x}} + \Gamma \mathbf{ \dot{x} } - \mathbf{f} \|^2_{\Gamma^{-1}} dt
\end{align}
and
\begin{align}
\label{over_func}
O[\mathbf{x} ] &= \int_0^T \| \mathbf{ \dot{x} } - \Gamma^{-1} \mathbf{f} \|^2_{\Gamma} dt
\end{align}
in the limit $\epsilon \rightarrow 0$, respectively (see e.g., \cite{F_W, ld_tech}). 

The minimizers of these functionals are often called instantons. They give a higher order correction to the leading order approximation of noiseless dynamics, taking into account the long-term effects of small noise.

Of particular interest is when $\mathbf{f}$ is the gradient of a potential, i.e.  $\mathbf{f} = -\nabla U$ for a function $U: \mathbb{R}^n \rightarrow \mathbb{R}$. Suppose one is interested in the transition from a local minimum $\mathbf{x}_A$ of $U$ to a saddle point $\mathbf{x}_S$. If there exists a heteroclinic connection from $\mathbf{x}_S$ to $\mathbf{x}_A$ in the noiseless overdamped system as well as a heteroclinic connection in a modification of the noiseless Langevin system, then the values of the functionals \ref{lang_func} and \ref{over_func} coincide; see \sref{prob_form_sec}. Furthermore, the existence of a heteroclinic connection in the former system does not guarantee the existence for a counterpart in the latter. As we will see in \sref{simple_sec}, if the friction coefficient is sufficiently low, the correspondence between these two heteroclinic connections can be broken. This implies that the minimal values of the two functionals no longer need to coincide. More generally, if the friction coefficient matrix has at least one sufficiently low eigenvalue, this possibility exists.

Consequently, the minimizing paths themselves can also be markedly different between the overdamped limit and the Langevin equation. This is no surprise given that the overdamped limit of the Langevin equation is a reversible diffusion process governed by a gradient system whereas the inertial Langevin system is irreversible, non-gradient, and non-uniformly-elliptic. We will describe how to compute the Langevin instantons in \sref{num_method}, based on structures of the functionals \ref{lang_func} and \ref{over_func} illustrated in \sref{prob_form_sec}. It is oftentimes possible to leverage computations from the overdamped equation to obtain instantons for the Langevin functional, and the string method \cite{PhysRevB.66.052301, simplified_string} originally designed for overdamped Langevin only needs to undergo a slight change to adapt to a matrix-valued friction coefficient that depends on position and velocity. However, for reasons just mentioned, it is still possible to miss local minimizers of the action \ref{lang_func}, and when this happens, we resort to the first-order optimality condition, which is only utilized in \sref{sec:underdamped} for understanding the differences between underdamped and overdamped Langevin equations.

\sref{sec:fric_examples} discusses the details of the Langevin transition problem in cases, with increasing complexity based on different kinds of positive definite friction coefficient matrices: 
\begin{enumerate}
\item \textbf{Constant Overdamped}. When the eigenvalues of the friction coefficient are large, the overdamped limit and the Langevin equation bare many similarities. This case is well-known and well studied (see for instance, \cite{TITULAER1978321}, \cite{dynamical_theory}, \cite{handbook_stoch}, \cite{stochastic_proc}). \sref{sec:overdamped} provides a concise review.

\item  \textbf{Constant Underdamped}. When the eigenvalues of the friction coefficient matrix are sufficiently low,  new phenomena distinctive from the overdamped limit can arise in the Langevin system. Physically this case may be motivated by considering situations where inertial effects are important. Fundamental differences appear when heteroclinic connections in the overdamped limit have no analog in the corresponding Langevin equation. Additionally, it is no longer the case that the global minimizer of the Freidlin-Wentzell action is a ``time-reversed" trajectory. We first illustrate this using a standard 4-dimensional test problem and then provide an explanation based on semi-analytical understanding of a 2-dimensional example in \sref{sec:underdamped}. Note that there have been great work on friction approaching zero asymptotics (e.g., \cite{stochastic_proc,frictionless_limit}), and the scope of this article is complementary as we focus on finite friction coefficients and masses.

\item \textbf{Position Dependent}. In addition we analyze position dependent friction coefficient matrices. See \cite{small_mass_state_dependent} and references therein for why such frictions are worth studying. Another motivation for position-dependent friction is to model certain biological systems -- when interactions with a solvent are modeled by noise and friction, the hydrophobic and hydrophilic regions of lipid bilayer, for instance, correspond to different friction coefficients \cite{position_dependent_friction_bio}.

To illustrate what difference can be induced by the position   dependence, we give an example in \sref{sec:position_dependent}, where we see that a localized change in the friction coefficient at a critical location can lead to a global change in the instanton. Two different definitions of an overdamped `limit' will also be discussed --- one preserves the invariant distribution of Boltzmann-Gibbs, and the other is consistent in terms of rare events.

\item \textbf{Velocity Dependent}: The structure of the Langevin functional also allows for the consideration of (both position and) velocity dependent friction coefficients, even though there is no more overdamped limit and Boltzmann-Gibbs is in general no longer an invariant distribution either. Velocity dependent friction coefficient can be viewed as a more flexible model of friction allowing for deviations from linear models \cite{nonlinear_friction_nature,OLSSON1998176}. The long term goal is to be able to address more complicated frictions such as those used to model swarming (e.g., \cite{levine2000self, d2006self}) and astrophysical dissipation (e.g., \cite{eggleton1998equilibrium, mardling2002calculating}), but investigations in this article are restricted to cases when zero velocity corresponds to stable fixed points or saddles.

To correctly compute instantons in the presence of velocity dependent friction coefficient (i.e., friction nonlinear in velocity), modifications need to be made. In particular, if the friction is not symmetric with respect to sign change in velocity, one has to consider \textbf{two} Langevin-type systems. See \sref{sec:string_method} and \sref{sec:momentum_dependent}.
\end{enumerate}

There are many great surveys of how to find minimizing trajectories in dynamical systems with noise, and here we provide an incomplete list of two most recent ones, \cite{Grafke2017} and \cite{Forgoston_2017}. The Langevin system considered here may be understood formally via a transverse decomposition of a first-order system (see \cite{Bouchet2016} and additionally \cite{MTao2017}), however the noise is degenerate (i.e., non-uniformly-elliptic diffusion). Versatile numerical methods  based on action minimization such as \cite{CPA:CPA20005, doi:10.1063/1.2830717, WAN20118669, CPA:CPA20238, VaHe08b, HeVa08c, lindley2013iterative, instanton_in_fluids, Grafke2017} work for nongradient systems and infinite dimensional systems too, but their applications to our system require adaptations, also due to the degenerate noise. Powerful theories based on hypoellipticity \cite{rogers1994diffusions,nualart2006malliavin} and hypocoercivity \cite{dric2009hypocoercivity} do provide tools for analyzing the degenerate and irreversible (but not too irreversible) system of Langevin, but our simple theoretical derivation will not require them.

\section{Problem Formulation}
\label{prob_form_sec}
In order to simplify analysis of the Langevin equation, we first change to mass coordinates, as in \cite{string_in_coll},  $X = M^{-1/2} Q$, $V = M^{-1/2} W$, $U(X) = Z(Q)$, $\frac{\partial Q}{\partial X} = M^{1/2}$,  so that 
\begin{equation}
\label{lang2}
\begin{array}{r@{}l}
 d X &= V dt    \\
 M d V &= -\Gamma  V  dt - \partial_X U dt+  \epsilon  \Gamma^{1/2} d W
\end{array}
\end{equation}
becomes
\begin{align}
\begin{array}{r@{}l}
\label{mass_coord}
 d Q &= W dt  \\
 d W &= - M^{-1/2} \Gamma M^{-1/2} W  dt - \partial_Q Z dt+  \epsilon  M^{-1/2} \Gamma^{1/2} d W .
\end{array}
\end{align}
The correlation matrix is $M^{-1/2} \Gamma^{1/2} (M^{-1/2} \Gamma^{1/2} )^T   = M^{-1/2} \Gamma M^{-1/2}$, thus we may effectively replace this system  by  
\begin{align}
\label{mass_coord2}
\begin{array}{r@{}l}
 d Q &= W dt  \\
 d W &= - M^{-1/2} \Gamma M^{-1/2} W  dt - \partial_Q Z dt+  \epsilon  \left( M^{-1/2} \Gamma M^{-1/2} \right)^{1/2} d W
 \end{array}
\end{align}
without changing relevant statistics. In these coordinates we have a system of identity mass with effective friction coefficient $M^{-1/2} \Gamma M^{-1/2}$. In this regard low mass and high friction are similar.  Thus, with this rescaling in mind, we take mass to be the identity matrix from now on.

Although the global minimizers of functionals \ref{lang_func} and \ref{over_func} are the only ones relevant for the most likely transitions, we relax this requirement and compute local minimizers. The rationale is that local minimizers are often associated to dynamical structures such as saddle points (see \eqref{eq_actionFull} and \eqref{eq_actionOverdamped} and discussions that follow), and if one can exhaust these structures (this is often easier), all local minimizers and hence the global one, can be found. We are interested in the case where the position boundary conditions coincide for the minimization of the overdamped and Langevin functionals, $\mathbf{x}(0) = \mathbf{x}_A$, $\mathbf{x}(T) = \mathbf{x}_B$, and the Langevin functional minimization is augmented with additional velocity boundary conditions, $\dot{\mathbf{x}}(0) = \mathbf{v_A}$, $\dot{\mathbf{x}}(T) = \mathbf{v_B}$. We typically choose homogeneous velocity boundary conditions, $\mathbf{v_A} = \mathbf{v_B} = \mathbf{0}$.

Utilizing the identity $\|a+b\|^2=\|a-b\|^2+4\langle a, b\rangle$ we rewrite the functionals \ref{lang_func} and \ref{over_func} as follows,
\begin{align}
L_T[\mathbf{x} ] &= \int_0^T \| \ddot{\mathbf{x}} + \Gamma \dot{ \mathbf{x} } + \nabla U \|_{\Gamma^{-1}}^2  dt \nonumber\\
&=\int_0^T \| \ddot{\mathbf{x}} - \Gamma \dot{ \mathbf{x} } + \nabla U \|_{\Gamma^{-1}}^2  dt + 4 \dot{\mathbf{x}}^T \left( \ddot{\mathbf{x}} + \nabla U \right) dt \nonumber\\
&= \int_0^T \| \ddot{\mathbf{x}} - \Gamma \dot{ \mathbf{x} } + \nabla U  \|_{\Gamma^{-1}}^2  dt + 4 \left[ \frac{1}{2}\| \mathbf{v_B} \|^2 - \frac{1}{2} \| \mathbf{v_A}\|^2 + U\left( \mathbf{x_B}\right)  - U \left( \mathbf{x_A}\right)  \right]	\label{eq_actionFull}
\end{align}
and 
\begin{align}
O_T[\mathbf{x} ] &= \int_0^T \| \mathbf{ \dot{x }} + \Gamma^{-1} \nabla U  \|_{\Gamma}^2 t \nonumber\\
&= \int_0^T \| \mathbf{ \dot{x }} - \Gamma^{-1} \nabla U  \|_{\Gamma }^2  + 4 \dot{\mathbf{x} }^T \nabla U  dt \nonumber\\
&= \int_0^T  \| \mathbf{ \dot{x }} - \Gamma^{-1} \nabla U  \|_{\Gamma }^2  dt    + 4 \left[ U\left( \mathbf{x_B}\right)  - U \left( \mathbf{x_A}\right)  \right] .	\label{eq_actionOverdamped}
\end{align}
These calculations show that the functionals are bounded below, i.e.
\begin{align}
\label{lower_lang}
L_T[\mathbf{x} ] &\geq   \max \{ 4 \left[ \frac{1}{2}\| \mathbf{v_B} \|^2 - \frac{1}{2} \| \mathbf{v_A}\|^2 + U\left( \mathbf{x_B}\right)  - U \left( \mathbf{x_A}\right)  \right] , 0 \}\\
\label{lower_over}
O_T[\mathbf{x} ] &\geq \max \{ 4 \left[ U\left( \mathbf{x_B}\right)  - U \left( \mathbf{x_A}\right)  \right], 0 \}
\end{align}
and in particular the lower bounds coincide if $ \| \mathbf{v_B} \|^2 -  \| \mathbf{v_A}\|^2 = 0$. 

The two main cases that we focus on here are
\begin{itemize}
\item The transition from a saddle point $\mathbf{x}_S$ to a local minimum $\mathbf{x}_B$ of $U$, with velocity boundary conditions $\mathbf{v_A} = \mathbf{v_B} = 0$, abbreviated $\mathbf{x_S} \rightarrow \mathbf{x_B}$.
\item The transition from a local minimum $\mathbf{x}_A$ to a saddle point $\mathbf{x}_S$ of $U$, with velocity boundary conditions $\mathbf{v_A} = \mathbf{v_B} = 0$, abbreviated $\mathbf{x_A} \rightarrow \mathbf{x_S}$.
\end{itemize}
The velocity and position boundary conditions guarantee that we are considering transitions between fixed points in both the Langevin and overdamped case. From this point on we consider the $T \rightarrow \infty$ limit. 

In the former case, if there exists a heteroclinic connection in the deterministic dynamics (for both systems) from $\mathbf{x}_S$ to $\mathbf{x}_B$ then the minimum action is $0$, since the minimizer is exactly given by deterministic dynamics (in the infinite time limit), i.e.
\begin{align}
\label{det_lang}
\ddot{\mathbf{x}} + \Gamma \dot{ \mathbf{x} } + \nabla U  &= 0 \\
\label{det_over}
 \dot{ \mathbf{x} } +  \Gamma^{-1} \nabla U  &= 0.
\end{align} 
However, it is possible for heteroclinic connections that exist in \ref{det_over} to not exist in \ref{det_lang} and vice versa as we show in \sref{simple_sec}.

For $\mathbf{x_A} \rightarrow \mathbf{x_S}$ it is also possible to find the minimizing trajectory which achieves equality in \ref{lower_lang} and \ref{lower_over}. We first focus on the case where $\Gamma$ depends only on position and assume that there exists a heteroclinic connection from $\mathbf{x_A}$ to $\mathbf{x_S}$ in the following systems:
\begin{align}
\label{det_lang2}
\ddot{\mathbf{x}} - \Gamma(\mathbf{x}) \dot{ \mathbf{x} } + \nabla U  &= 0 \\
\label{det_over2}
 \dot{ \mathbf{x} } -  \Gamma^{-1}(\mathbf{x} ) \nabla U  &= 0
\end{align}
In this case, these connections are the action minimizing trajectories. Reversing time $\tau = T - t$, and denoting derivatives with respective to the new variable with $'$, i.e. $\dot{x} = -x ' $, we see
\begin{align}
\label{rdet_lang2}
\mathbf{x} ''  + \Gamma(\mathbf{x}) \mathbf{x}' + \nabla U(\mathbf{x})  &= 0 \\
\label{rdet_over2}
 \mathbf{x}'  + \Gamma^{-1}(\mathbf{x}) \nabla U(\mathbf{x})   &= 0
\end{align} 
which are exactly the same as \ref{det_lang} and \ref{det_over} with respect to the new variables. The minimizing trajectory becomes the same as the noiseless trajectory from $\mathbf{x_S} \rightarrow \mathbf{x_A}$ after a sign change of the velocity variable in the Langevin case.

If the friction coefficient matrix depends on velocity then the minimizing trajectory must satisfy
\begin{align}
\label{det_lang3}
\mathbf{x''}  + \Gamma(\mathbf{x} , - \mathbf{x}') \mathbf{x}' + \nabla U(\mathbf{x})  &= 0 
\end{align}
which only coincides with \ref{det_lang} if $\Gamma(\mathbf{x} , - \mathbf{x}')  = \Gamma(\mathbf{x} , \mathbf{x}') $. We call \ref{det_lang3} the time-reversed Langevin equation. Heteroclinic connections that exist in \ref{det_lang2} do not necessarily exist in \ref{det_lang3}.

If there is no heteroclinic connection between points of phase space, the time-reversed dynamics do not correspond to minimizers. To understand this case better, we directly use the Euler-Lagrange equation of the variational principle (note: the appearance of $\mathbf{\ddot{x}}$ normally would require the introduction of a jet bundle instead of the standard tangent bundle, but there is no need to concern this technicality because we work with flat position space $\mathbb{R}^n$):
\begin{align}
A[ \mathbf{x} ] &= \int_0^T \mathcal{L}(\mathbf{x}, \mathbf{\dot{x}} , \mathbf{\ddot{x} } ) dt, \\
\frac{\delta A}{\delta \mathbf{x} } &= \frac{\partial \mathcal{L}}{\partial \mathbf{x} }  - \frac{d}{dt} \frac{\partial \mathcal{L}}{\partial \mathbf{\dot{x}} } 
+ \frac{d^2}{dt^2}\frac{\partial \mathcal{L}}{\partial \mathbf{\ddot{x}} } = 0.
\label{eq_EL}
\end{align}
If the friction coefficient matrix is isotropic $\Gamma = \gamma \mathbb{I}$ then simplifications occur for the Langevin functional \ref{lang_func} and the Euler-Lagrange equation becomes
\begin{align}
\frac{d^4}{dt^4}\mathbf{x} - \frac{d^2}{dt^2}\mathbf{f} - \gamma^2 \frac{d^2}{dt^2} \mathbf{x}  - \nabla \mathbf{f} \frac{d^2}{dt^2} \mathbf{x}  + (\nabla \mathbf{f} )^T  \mathbf{f} = 0.
\label{eq_ELspecific}
\end{align}
We solve this boundary value problem numerically: \sref{sec:eul_lag} uses gradient descent, a finite time-horizon approximation, and pseudospectral discretization.

\section{Numerical Methods}
\label{num_method}
To compute minimizers to the Freidlin-Wentzell action \ref{lang_func}, we utilize three different methods: the string method, the numerical integration of dynamics, and directly solving the boundary value problem.  We review each method and show how to implement them in the following subsections. Their use in practice is illustrated in \sref{sec:fric_examples}.

\subsection{String method}
\label{sec:string_method}
The string method is both a way of locating saddles between local minima as well as a way to approximate local minimizers of the overdamped Freidlin-Wentzell action  \cite{PhysRevB.66.052301}, \cite{simplified_string}.
To utilize this method one proceeds as follows:
\begin{enumerate}
\item  First define an initial ``string", i.e, a sequence of points from one local minimum to another, or simply a sequence of points that lie across a separatrix. For example, in many cases one can choose a straight line between one local minimum and another.
\item Evolve each point on the string according to the deterministic dynamics for a single time-step.
\item Calculate the total length of the string and then evenly redistribute the set of points on the string.
\item Repeat 2-3 until convergence.
\end{enumerate}
The path interpolated by the sequence of points (also called images) is the ``string".  

The string method can also be applied to the Langevin system by writing the equation as a first order system. Furthermore it can approximate a local minimizer of the corresponding Langevin Freidlin-Wentzell action by making a few adaptations. As stated in the previous section, once a saddle is located, the minimizer of the Freidlin-Wentzell action from a local minimum to a saddle obeys time-reversed dynamics, i.e.
\begin{align}
\begin{array}{r@{}l}
\dot{\mathbf{x} } &= \mathbf{v}  \\
\dot{\mathbf{v} } &= - \nabla U + \Gamma( \mathbf{x} , \mathbf{v} ) \mathbf{v}
\end{array}
\end{align}
as long as a heteroclinic connection exists from the minimum to the saddle. Reversing time, and making the change of variables $\mathbf{v} = - \mathbf{w}$, yields the following system
\begin{align}
\label{time_reversed_langevin}
\begin{array}{r@{}l}
\mathbf{x}'  &= \mathbf{w}   \\
\mathbf{w}' &= - \nabla U - \Gamma( \mathbf{x} , -\mathbf{w} ) \mathbf{w}
\end{array}
\end{align}
where $\mathbf{x} ' = - \dot{ \mathbf{x} }$. This is the noiseless Langevin equation, but with friction coefficient matrix $\Gamma(\mathbf{x}, - \mathbf{w})$ as opposed to $\Gamma(\mathbf{x}, \mathbf{w})$. If the friction coefficient matrix is an even function of the velocity then one can run the string method without modification to compute the correct positions $\mathbf{x}$ of the string. Once the string has converged, the velocity terms need to undergo a sign flip in the ``uphill" part of the string.

\begin{figure}
\begin{center}
\includegraphics[width=1.0\textwidth]{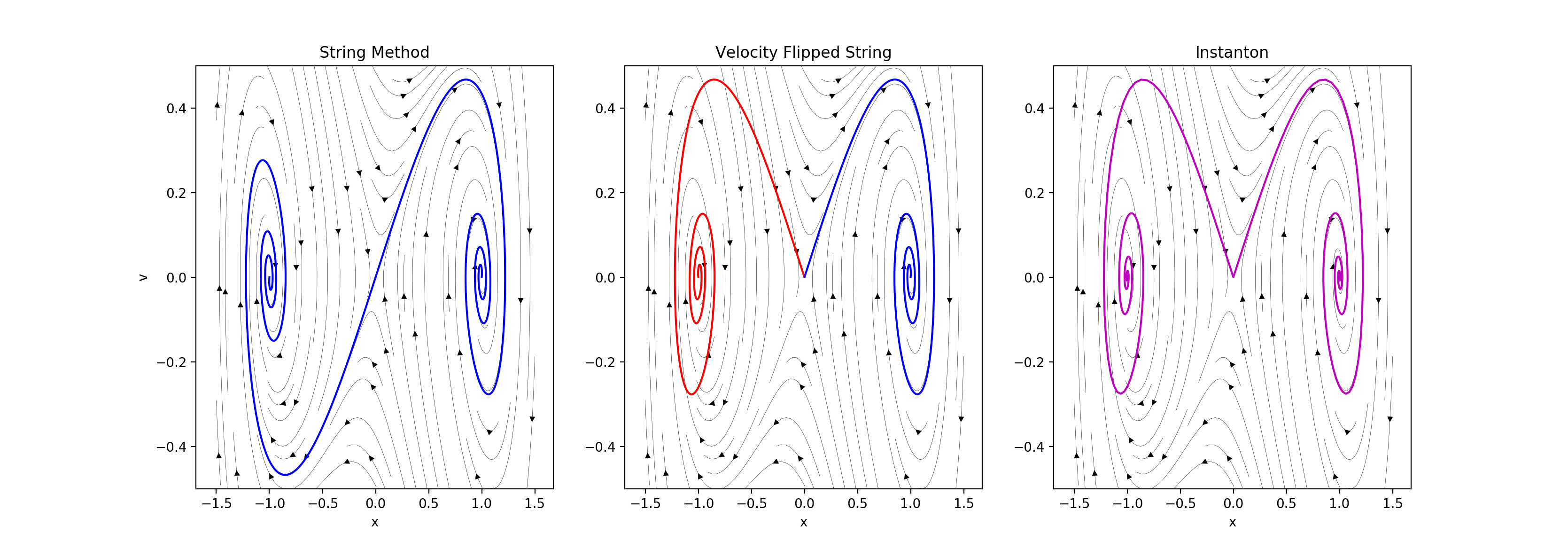}
\caption{[Color online] Phase portrait of the double well potential utilizing the string method (left), changing the sign on velocity in the ``uphill" transition (middle), and computing the solution to the Euler-Lagrange equations (right). The middle and right panels are minimizers of the Freidlin-Wentzell action for the $(-1,0)$ to $(1,0)$ transition. The streamlines correspond to dynamical trajectories.}
\label{fig:double_well}
\end{center}
\end{figure}

We illustrate what we mean by the sign flip in \fref{fig:double_well}. Here we are computing the transition path in the double well potential 
\begin{align}
U(x) &= \frac{1}{4}(1-x^2)^2
\end{align}
from $(x_A, v_A) = (-1, 0)$ to $(x_B, v_B) = (1, 0)$ for $\gamma = 0.5$. There is a saddle at the point $(0,0)$ and it serves as the partition from ``uphill" to ``downhill". The true action minimizing path may be viewed as going ``uphill" from $(-1,0)$ to $(0,0)$ and then proceeding ``downhill" along dynamics from $(0,0)$ to $(1,0)$. The left most figure is obtained by evolving a string; however, this is not the action minimizing path. Along the uphill part, from $(-1,0)$ to $(0,0)$, the velocity variable needs to be multiplied by a negative sign to produce the true action minimizing path. Doing so the middle figure is obtained. The rightmost figure is the action minimizing path obtained as a solution to the Euler-Lagrange equations \ref{eq_ELspecific}, via the methods of \sref{sec:eul_lag}.

\begin{figure}
\begin{center}
\includegraphics[width=1.0\textwidth]{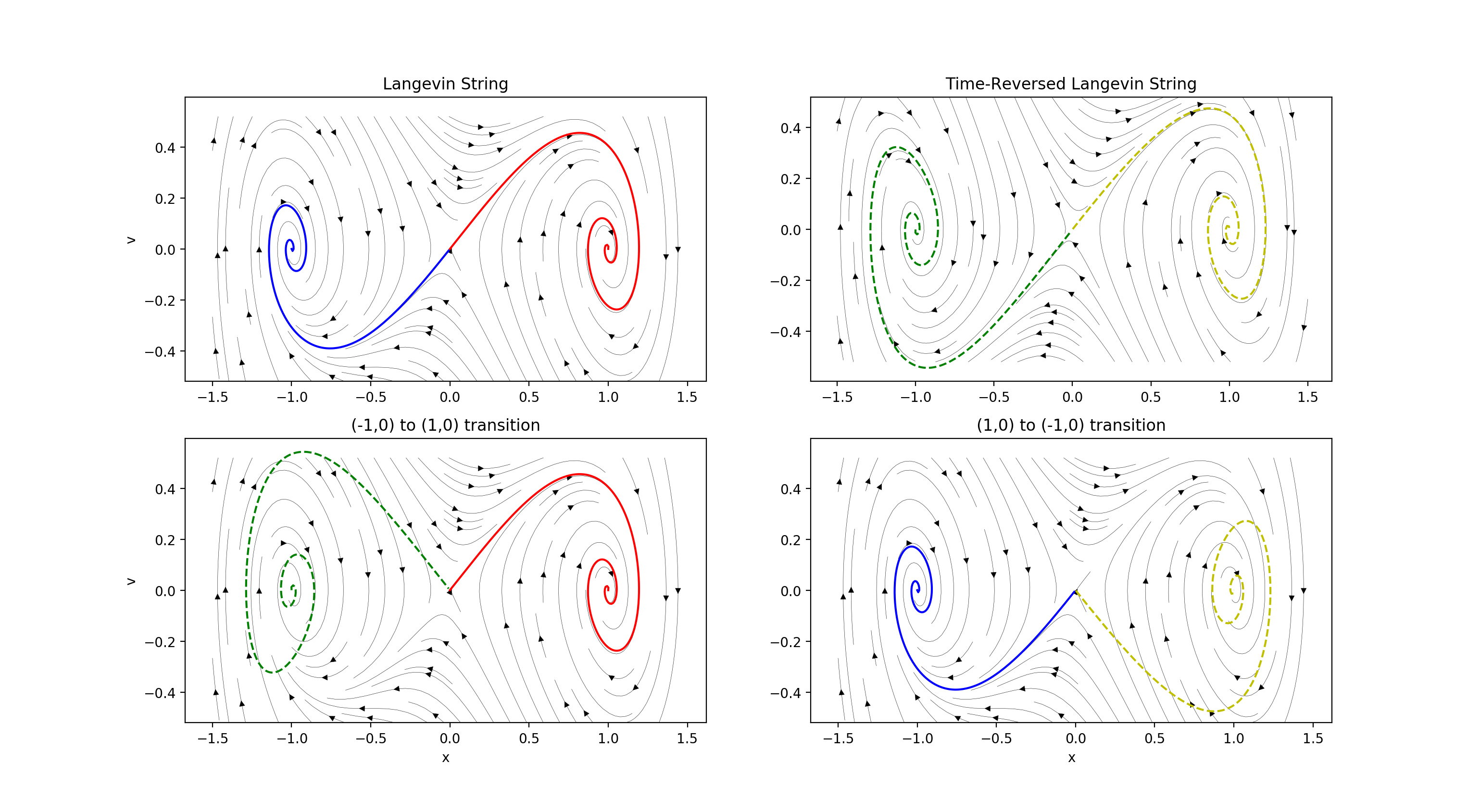}
\caption{[Color online] Phase portrait of the double well potential with nonlinear friction. The dynamical trajectories of the Langevin system (top left) and the time-reversed Langevin system (top right) are calculated via the string method. The $(-1,0)$ to $(1,0)$ transition (bottom left) and $(1,0)$ to $(-1,0)$ transition (bottom right) are constructed from the strings of the top panels. The streamlines correspond to the Langevin system for all but the top right panel. The top right panel has the stream lines of the time-reversed Langevin system.}
\label{fig:nonlin_double_well}
\end{center}
\end{figure}

 If $\Gamma(\mathbf{x} ,- \mathbf{w}) \neq \Gamma(\mathbf{x} , \mathbf{w})$, the procedure needs to be modified. We illustrate this with the following (noiseless) system:
 \begin{align}
 \label{first_system}
 \begin{array}{r@{}l}
 \dot{x} &= v \\
 \dot{v} &= - 0.5 e^{v (x-0.5)}v - x - x^3.
 \end{array}
 \end{align}
 Here the nonlinear friction coefficient is $\Gamma(x,v) = 0.5 e^{v (x-0.5)}$.
 The easiest solution is to evolve \textbf{two} strings. The first obeys the noiseless Langevin dynamics (as in \ref{first_system}) and the latter obeys \ref{time_reversed_langevin}. With regards to \ref{first_system} the corresponding time-reversed system is 
 \begin{align}
 \label{second_system}
 \begin{array}{r@{}l}
 \dot{x} &= v \\
 \dot{v} &= - 0.5 e^{-v (x-0.5)}v - x - x^3,
 \end{array}
 \end{align}
 where the the nonlinear friction coefficient is $\Gamma(x,-v) = 0.5 e^{-v (x-0.5)}$.
  We call this system the time-reversed (noiseless) Langevin equation. The instanton is then a concatenation of different halves (separated by the saddle point) of these two different strings. For the uphill part one takes the string for the time-reversed Langevin equation with a sign flip on velocity and for the downhill part one takes the Langevin string.  For example, in \fref{fig:nonlin_double_well} the two systems, \ref{first_system} and \ref{second_system}, are simulated via the string method in the top two panels. The instanton for the system for two different transitions--$(-1,0)$ to $(1,0)$ and $(1,0)$ to $(-1,0)$--are shown in the bottom two panels. The $(-1,0)$ to $(1,0)$ transition is constructed by taking the $(0,0)$ to $(1,0)$ half of the string corresponding to \ref{first_system} (downhill dynamics), and concatenating it with the $(-1,0)$ to $(0,0)$ half of the time-reversed system with velocity multiplied by a negative sign (uphill trajectory). An analogous procedure is used to construct the $(1,0)$ to $(-1,0)$ transition.
  
 It is important to keep in mind that the string method can only compute a local minimizer of the Freidlin-Wentzell action, and different initializations of the string could give rise to different local minimizers. On a practical level, one can try different string initializations and choose the final string that best minimizes the Freidlin-Wentzell action.

\subsection{Saddle points and dynamics}
\label{sec:saddle_points}
 The Langevin instantons are generally more costly to compute than the overdamped case as they can be more oscillatory. For instance, the method described in \sref{sec:string_method} often under-resolves the instanton unless it is discretized by a large number of points. To reduce the computational cost, knowledge of the overdamped dynamics (when they exist) can  be exploited. Such knowledge includes, in particular, the saddle point and the perturbation directions off that saddle to the local minima. A sufficiently resolved string produces the minimizing path for the overdamped system, which includes the approximate location of a saddle between two minima as well as the perturbation directions. We mention that if one is purely interested in just the saddle, the computation of the string can be avoided, see for example \cite{gentlest_ascent2011, ren2013climbing}.

When both heteroclinic connections exist (see \sref{prob_form_sec}), we utilize these approximate perturbation directions of the overdamped system to obtain approximate perturbation directions of the Langevin system and run dynamics to compute the instanton.

This algorithm proceeds as follows:
\begin{enumerate}
\item Calculate the overdamped string.
\item Locate the saddle(s), and then utilize perturbations away from the saddle (as calculated by the string) to obtain perturbation directions for the inertial Langevin system.
\item Simulate the dynamics of the corresponding noiseless Langevin equations. Concatenate to construct the instanton.
\end{enumerate}
One way to locate a saddle point on a string is to evaluate the norm of the force at each point on the string and choose the one with $|\nabla U \left( x \right) |^2 \leq \epsilon$. If heteroclinic connections exist, it is possible to explicitly calculate the correct perturbation direction off the saddle for the Langevin system based on the overdamped result, see \sref{perturb_direc}. Another possibility is to compute underresolved Langevin strings (i.e. those in \sref{sec:string_method}) for approximating the perturbation directions.

\subsection{Euler-Lagrange equation}
\label{sec:eul_lag}
It is sometimes necessary to compute instantons directly by numerically solving the Euler-Lagrange equation. This is based on an iterative minimization of the Freidlin-Wentzell action analogous to existing results. In fact, great methods have been developed to compute local minima of \ref{over_func} which can be nongradient. Among them include the minimum action method \cite{CPA:CPA20005}, adaptive minimum action method \cite{doi:10.1063/1.2830717} and its higher-order version \cite{WAN20118669}, the geometric minimum action method \cite{CPA:CPA20238, VaHe08b, HeVa08c}, and methods based off of dynamic programing \cite{Dahiya2018}. It is also possible to utilize the Hamiltonian formulation, see \cite{CPA:CPA20238, Grafke2017, lindley2013iterative}. Note, though, with regards to the Langevin problem considered here, there is no need to use these more generic and more costly approaches, unless the $\mathbf{x_A} \rightarrow \mathbf{x_S}$ or $\mathbf{x_S} \rightarrow \mathbf{x_B}$  heteroclinic connection does not exist (whether they exist is verifiable by methods in \sref{sec:string_method} and \sref{sec:saddle_points}).

Action minimization methods have a natural extension to the Langevin setting \ref{lang_func}. For example, one can use gradient-descent and a finite $T$ approximation to compute minimizers, similar to what was originally done with the minimum action method \cite{CPA:CPA20005}. Due to the second derivatives arising in the functional, however, the resulting boundary value problem will be fourth-order. Specifically the Euler-Lagrange equations for the Langevin functional are 
\begin{align}
\frac{\delta L }{\delta \mathbf{x}} &= \frac{d^4}{dt^4}\mathbf{x} - \frac{d^2}{dt^2}\mathbf{f} - \gamma^2 \frac{d^2}{dt^2} \mathbf{x}  - \nabla \mathbf{f} \frac{d^2}{dt^2} \mathbf{x}  + (\nabla \mathbf{f} )^T  \mathbf{f} = 0
\end{align}
under the assumption that the friction coefficient matrix is isotropic, $\Gamma = \gamma \mathbb{I}$ and identity mass. 

As per usual with gradient-descent, we introduce pseudo-time $\tau$ and evolve
\begin{align}
\partial_\tau \mathbf{x}  &= -\frac{\delta L }{\delta \mathbf{x}}
\end{align}
forward in pseudo-time until $\frac{\delta L }{\delta \mathbf{x}}$ is sufficiently small. When discretizing pseudo-time we treat the linear terms implicitly in order to mitigate stiffness due to hyper-diffusion and diffusion, as is commonly done in numerical partial differential equations. Explicitly, the following linear update is solved at every time step
\begin{align} 
\label{four_bvp}
 \left( \frac{d^4}{dt^4} - \gamma^2 \frac{d^2}{dt^2}  + \frac{1}{\Delta \tau} \right)\mathbf{x}_{n+1} &= \frac{d^2}{dt^2}\mathbf{f}_n + \nabla \mathbf{f}_n \frac{d^2}{dt^2} \mathbf{x}_n - (\nabla \mathbf{f}_n )^T  \mathbf{f}_n + \frac{1}{\Delta \tau } \mathbf{x}_{n} 
\end{align}
where $\Delta \tau$ is the  pseudo-time step size. This is a linear boundary value problem for $\mathbf{x}_{n+1}$ at every time-step. 

 Amongst the many methods to discretize time $t$ we utilize a modern form of spectral integration as in \cite{Greengard_1991} and \cite{Viswanath2015159} . The main idea behind the method is to rewrite \ref{four_bvp} in integral form and represent the solution utilizing Chebyshev polynomials. Although derivative operators are dense with respect to Chebyshev collocation points, integral operators in spectral space are banded. This leads to efficient and numerically robust evaluation of solutions. Products and convolutions are computed utilizing the pseudo-spectral method \cite{Boyd}.

It is worth noting that one can also compute minimizers to the action
\begin{align}
\label{relaxed}
\int_0^T  \left(\lambda \| \mathbf{ \dot{x} } - \mathbf{ v  } \|^2 + \|  \mathbf{ \dot{v}  } +\gamma  \mathbf{v} - \mathbf{f} \|^2 \right) dt,
\end{align}
where $\lambda$ is a large parameter. The advantange of utilizing this functional is that one can use the same methods that are applied to systems with non-degenerate noise; however, solving \ref{relaxed} leads to a stiff system of equations, resulting in expensive computations. When heteroclinic connections exist, the action minimizer coincides with that of a non-stiff version,
\begin{align}
\int_0^T  \left(\| \mathbf{ \dot{x} } - \mathbf{ v  } \|^2 + \|  \mathbf{ \dot{v}  } +\gamma  \mathbf{v} - \mathbf{f} \|^2 \right) dt,
\end{align}
as $\| \mathbf{ \dot{x} } - \mathbf{v} \|^2$ can be made zero pointwise. However, this relaxation may not work if one of the two heteroclinic connections doesn't exist (although when they both exist there is less need to iteratively minimize the action).

\section{The Differences}
\label{sec:fric_examples}
In what follows we examine local minimizers of the Langevin action \ref{lang_func} with matrix-valued and possibly position and velocity dependent friction coefficient $\Gamma$ to highlight generic differences and similarities with its overdamped limit \ref{over_func} and isotropic friction coefficients. At the end of each section we give generic recommendations on numerical methods.
 
\subsection{Large friction coefficient}
\label{sec:overdamped}
The overdamped case has been studied extensively as previously mentioned, (e.g. \cite{TITULAER1978321}, \cite{dynamical_theory}, \cite{handbook_stoch}, \cite{stochastic_proc}). We also note that the closely related problem of the small mass limit --- which we saw in \sref{prob_form_sec} is related to high friction via a rescaling of coordinates --- has been analyzed, see e.g., \cite{small_mass, small_mass_state_dependent, small_mass_variable_friction, FREIDLIN201345}. To recall the phenomenology concretely, we use an illustration based on the Mueller potential
\begin{align}
U(x,y) &= \sum_{i=1}^4 A_i \exp \left( a_i (x-x_i)^2 + b_i (x-x_i)(y-y_i) + c_i (y-y_i)^2\right),
\end{align}
where the parameters $A_i, a_i, b_i , c_i,x_i, y_i$ for $i=1,..,4$ are chosen the same as \cite{simplified_string}. The minimizer here can be computed via the string method or utilizing dynamics.
 
\begin{figure}
\begin{center}
\includegraphics[width=1.0\textwidth]{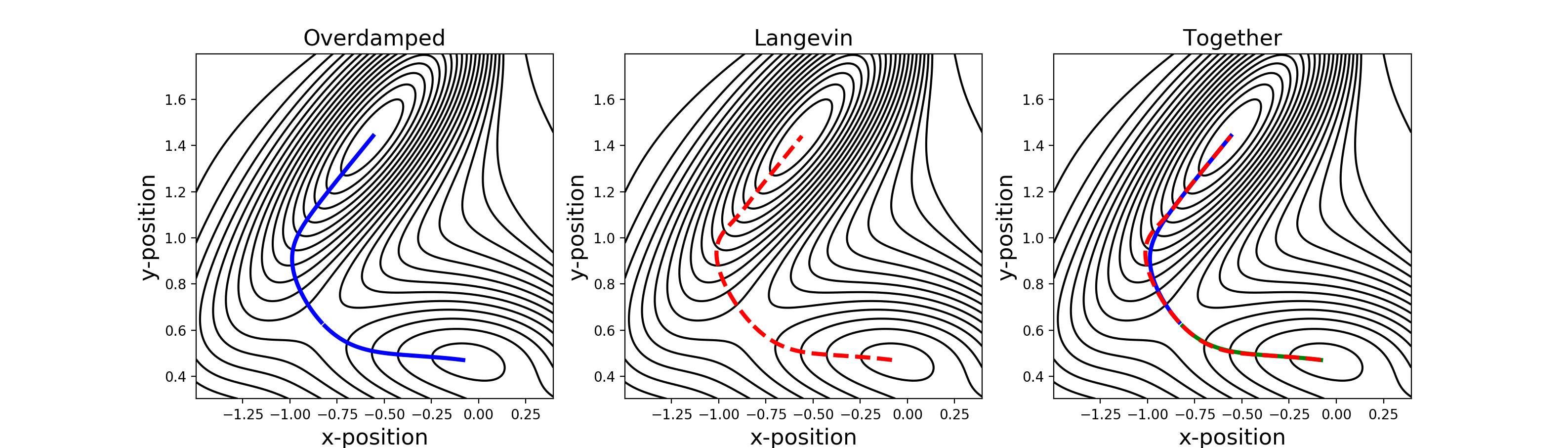}
\caption{[Color online] Instantons for the overdamped functional (left), the Langevin functional (middle), and both together (right). The primary difference between the overdamped trajectory and the Langevin trajectory is a slight overshoot in the upper half of the Langevin trajectory.}
\label{fig:overdamped}
\end{center}
\end{figure}

\fref{fig:overdamped} shows the trajectories of two local minimizers of the Freidlen-Wentzell action with an isotropic (i.e. scalar) friction coefficient $\gamma = 50$. We see that, after projecting to position space in the case of Langevin system, there is little difference in the trajectory of the Langevin equation and its overamped limit, except for a small amount of overshoot in the Langevin dynamics.
 
  In the high friction scenario this agreement is generic. If $\gamma$ is much larger than the spectral radius of the Hessian of $U$ in the domain of interest, the friction is high enough to justify the overdamped limit.
 
\paragraph{Recommendation} The general recommendation for this case is to focus on the overdamped limit, and the string method in its original form is a good choice for computing the instanton.

\subsection{Small constant scalar friction coefficient}
\label{sec:underdamped}
\label{simple_sec}
As we lower the friction coefficient, new situations may arise since inertial effects can now compete with the potential. We first illustrate this in the Mueller system and then provide a detailed analysis of a system with a sixth-order polynomial potential. The features mentioned here are generic and we have observed similar behavior in other systems.

\begin{figure}
\begin{center}
\includegraphics[width=1.0\textwidth]{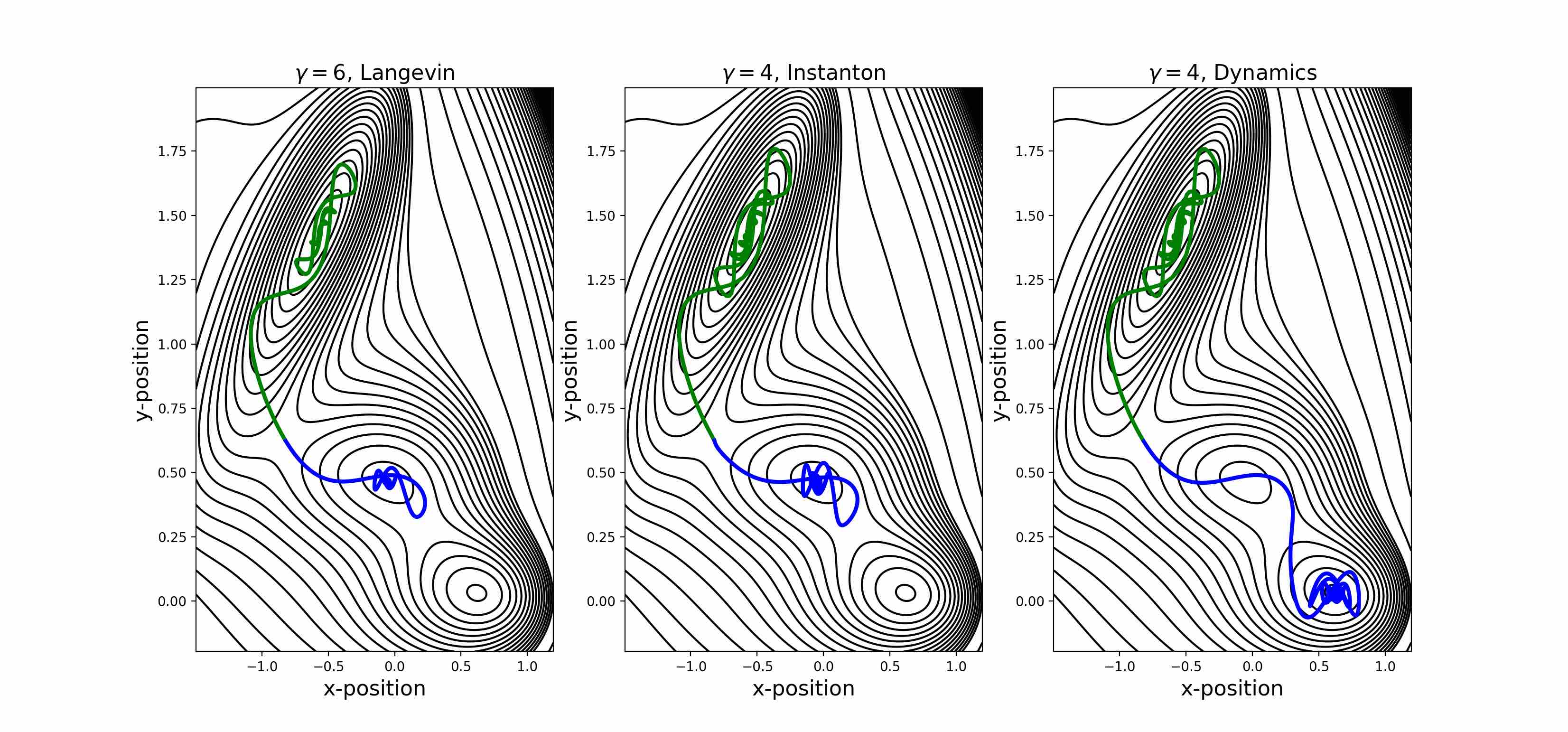}
\caption{[Color online] Dynamical trajectories and instanton of the Langevin Friedlin-Wentzell action for different friction coefficients in the underdamped regime. On the left ($\gamma = 6$) the dynamical trajectories settle to the same minima as the overdamped scenario and correspond to instantons of the Freidlin-Wentzell action. As we further decrease the friction coefficient to $\gamma = 4$ the instanton (middle) no longer coincides with a generic dynamical trajectory that starts from a perturbation off the saddle (right). }
\label{fig:mueller_skip}
\end{center}
\end{figure}

 We apply the method of \sref{sec:saddle_points} and \sref{sec:eul_lag} to various dissipation regimes of the Mueller system. When the friction becomes small but not too small, $\gamma = 6$ for instance in this case, oscillatory behaviors near minima are seen in the instanton (left panel of \fref{fig:mueller_skip}). These oscillations can be intuitively understood as an inertial effect due to the fact that the particle cannot stop or take sharp turns instantaneously, unless a lot of noise is used. However, the topology of the transition remains the same as the overdamped instanton.
 
Further lowering the friction coefficient, for instance to $\gamma = 4$, leads to a difference in the instanton and a dynamical trajectory. A generic dynamical trajectory settles to a different minimum, see the right panel of \fref{fig:mueller_skip}. In this case, the particle has enough momentum to overcome intermediary minimum and ends up in another minimum of the potential. One may think that this means a better perturbation direction needs to be chosen for a correct departure from the saddle point; however, in this case we posit that there no longer exists a heteroclinic connection between the saddle and the original target minimum.

\begin{figure}
\begin{center}
\includegraphics[width=0.5\textwidth]{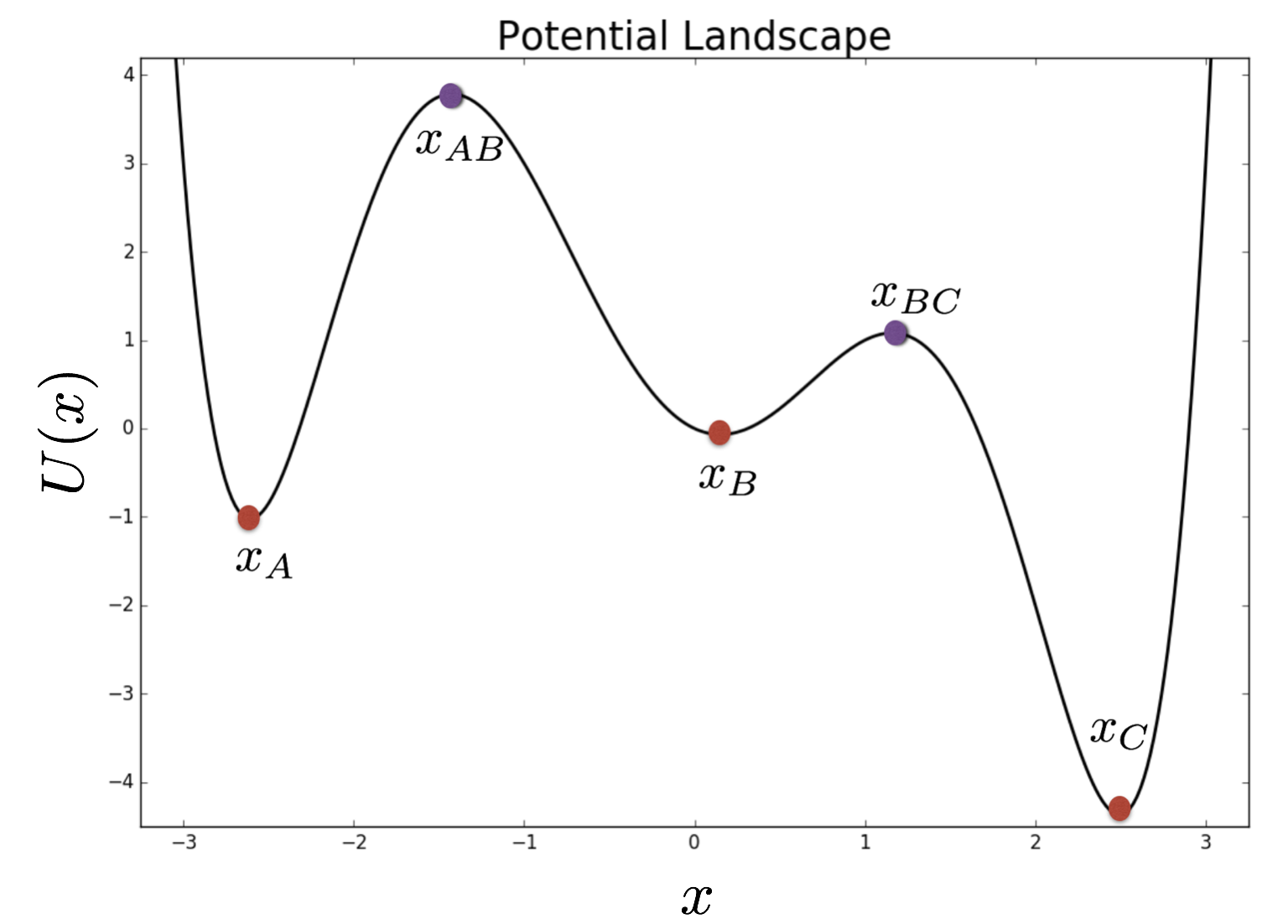}
\caption{[Color online] The sixth-order polynomial potential function. There are three local minima and two local maxima which correspond to saddle points in phase space.}
\label{fig:potential}
\end{center}
\end{figure}

To understand this possibility of losing heteroclinic connection, let us consider a simpler, one degree-of-freedom example (the governing SDE is therefore 2D ), where the potential is given by a sixth-order polynomial 
\begin{align}
U(x) &= \sum_{i=1}^6 a_i x^i .
\end{align}
$a_i \in \mathbb{R}$ for $i=1,...,6$ are such that there are three local minima at positions $x_A \leq x_B \leq x_C$ and saddles $x_{AB} \leq x_{BC}$. Furthermore $U(x_{ab}) \geq U (x_{bc} )\geq U(x_{b}) \geq U(x_{a}) \geq U(x_c)$. See \fref{fig:potential} for an example.

In the overdamped limit, the state space is 1-dimensional, and the transitions between different minima can only go through one path, i.e. the transition from $x_A \rightarrow x_C$ has to first pass through the minima at $x_B$. Furthermore the transitions must also pass through the saddles at $x_{AB} $ and $x_{AC}$.

In the Langevin system, however, this is not necessarily the case since the state space is 2-dimensional. More precisely, the transition path of $x_A \rightarrow x_B$ changes with the amount of friction in the system.  With a large amount of friction, the transition simply passes through the saddle $x_{AB}$ and slides into $x_B$ as a solution to \ref{det_lang}; however with less friction \ref{det_lang} may no longer have an infinite time solution with boundary conditions $x_{AB}$ and $x_B$.  Physically, this corresponds to the situation where the particle starts at the saddle $x_{AB}$ with infinitesimal velocity to the right, builds up enough momentum due to insufficient friction, so that after passing $x_B$ it overcomes the small barrier given by $x_{BC}$ and eventually settles at $x_C$. In this case there is no dynamical path from $x_{AB} \rightarrow x_B$, but there is a dynamical path from $x_{AB} \rightarrow x_C$. Note with even less friction it is still possible for the heteroclinic connection from $x_{AB}$ to $x_{B}$ to exist again (the particle can bounce back), but we focus on the case where this connection doesn't exist. Naturally, this phenomenon of potential-well-skipping will be pronounced if $U(x_{BC})-U(x_B) \ll U(x_{AB})-U(x_B) \ll U(x_{BC})-U(x_C)$.

There are three reasonable possibilities for the global action minimizer, when the boundary conditions are $x_{A}$ and $x_B$ (equipped with zero velocity) and yet there is no heteroclinic connection from $x_{AB}$ to $x_B$. The first corresponds to the dynamical paths / time-reversed paths $x_A \rightarrow x_{AB} \rightarrow x_{C} \rightarrow x_{BC} \rightarrow x_{B}$, each equipped with zero velocity. In this case, the local minimal action value of the Freidlin-Wentzell action is 
\begin{align}
A_{\text{loc}} &= 4 ( U(x_{AB} ) - U( x_A ) + U(x_{BC}) - U (x_C ) )
\end{align}
 The second possibility corresponds to $x_A \rightarrow x_{AB} \rightarrow x_{B}$, but the instanton from $x_{AB}$ to $x_B$ is no longer given by a simple deterministic dynamics. A more efficient minimizer of the action is obtained by deviating away from the deterministic path. The third possibility is that the minimizing path no longer transitions through the saddle $x_{AB}$ and instead transitions from $x_{A}$ to $x_{B}$ directly, without passing through any other critical points.
 
 Let us first compare the first two possibilities: in this example, it is more efficient to use noise to slow down the particle so that it settles at $x_B$. Naturally, this is the case if the difference $U(x_{BC}) - U(x_C)$ is sufficiently large, corresponding to a deep well at location $x_C$. The first possibility is not optimal as the particle needs a lot of noise to crawl back into the well of $x_B$, and it would rather spend a small amount of noise to slow down in the well of $x_B$ before getting trapped in the well of $x_C$. 
 
\begin{figure}
\begin{center}
\includegraphics[width=1.0\textwidth]{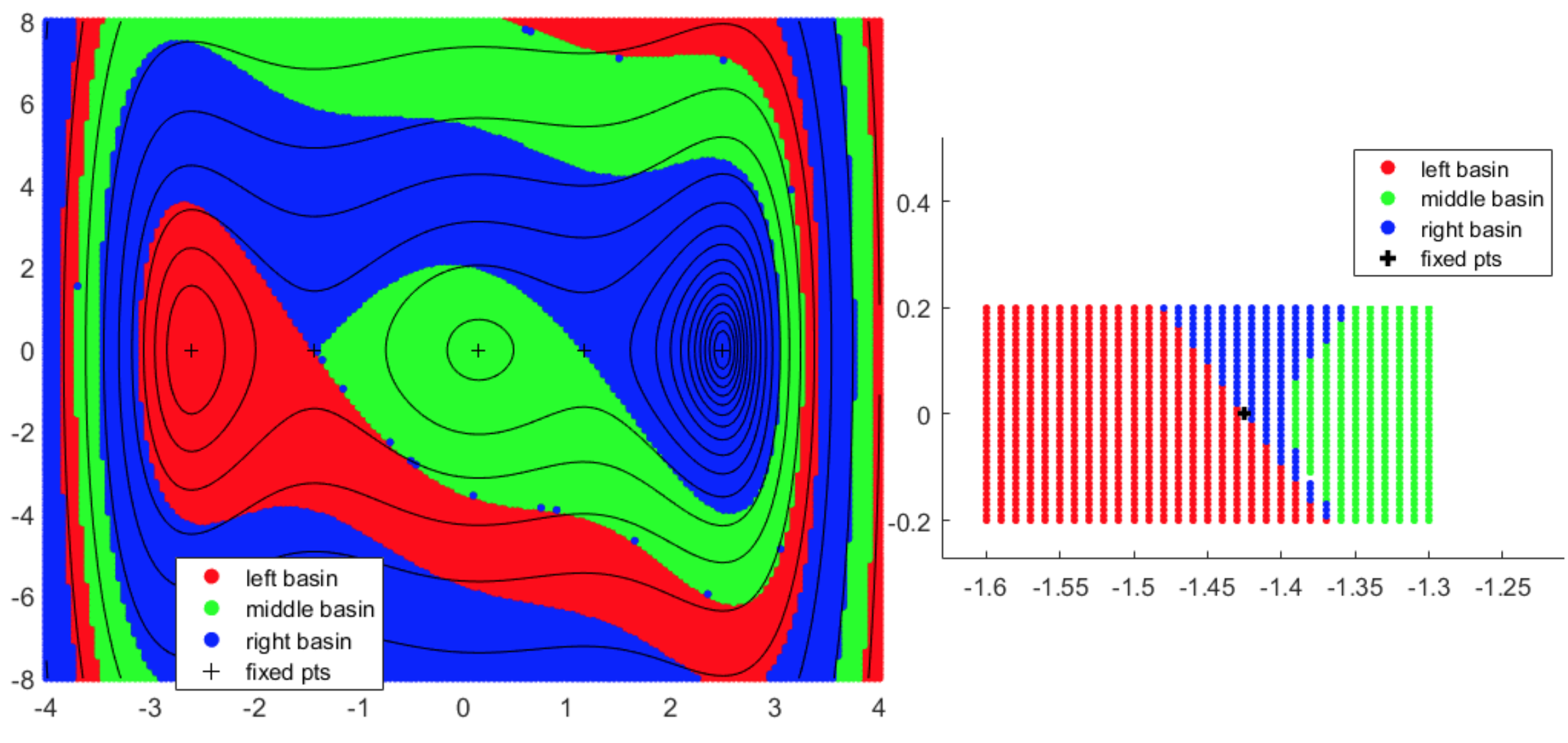}
\caption{[Color online] Basins of attraction for the Langevin system with the sixth order polynomial potential and $\gamma$ = 0.75.  }
\label{fig:basin}
\end{center}
\end{figure}

To be more specific, parameters used for producing \fref{fig:potential} are
\begin{align*}
[a_1, a_2, a_3,a_4,a_5,a_6] = \frac{1}{120}[-108,364,-15,-135,3,11 ],
\end{align*}
so that
\begin{align}
&x_{A} = -2.607291444942787 \text{, } x_{AB} = -1.4251363637044439 \text{, } x_{B} =  0.152393839371333 ,  \\
&x_{BC} =  1.1586416396791057 \text{, } x_{C} = 2.494119602324065
\end{align}
and
\begin{align*}
&U(x_{AB}) - U(x_A ) = 19.22792801 \text{, } U(x_{AB}) - U(x_B) = 15.41318608 \text{, } \\
&U(x_{BC} ) - U(x_B) = 4.59671282 \text{, } U(x_{BC}) - U(x_C) = 21.8046503
\end{align*}
In this case, when $\gamma = 1$ there exists a heteroclinic connection between $x_{AB}$ and $x_B$, whereas when $\gamma = 0.75$ there no longer exists a heteroclinic connection\footnote{The gamma for which this bifurcation first occurs appears is between $0.76$ and $0.75$.}. \fref{fig:basin} shows the basins of attraction for the $\gamma=0.75$ system, approximated by numerical simulations of a large amount of initial conditions. Zooming in close to $x_{AB}$ one sees that the basin of attraction for $x_C$ had been eroded away, cutting off any noiseless transition from $x_{AB}$ to $x_{B}$.

\begin{figure}
\begin{center}
\includegraphics[width=1.0\textwidth]{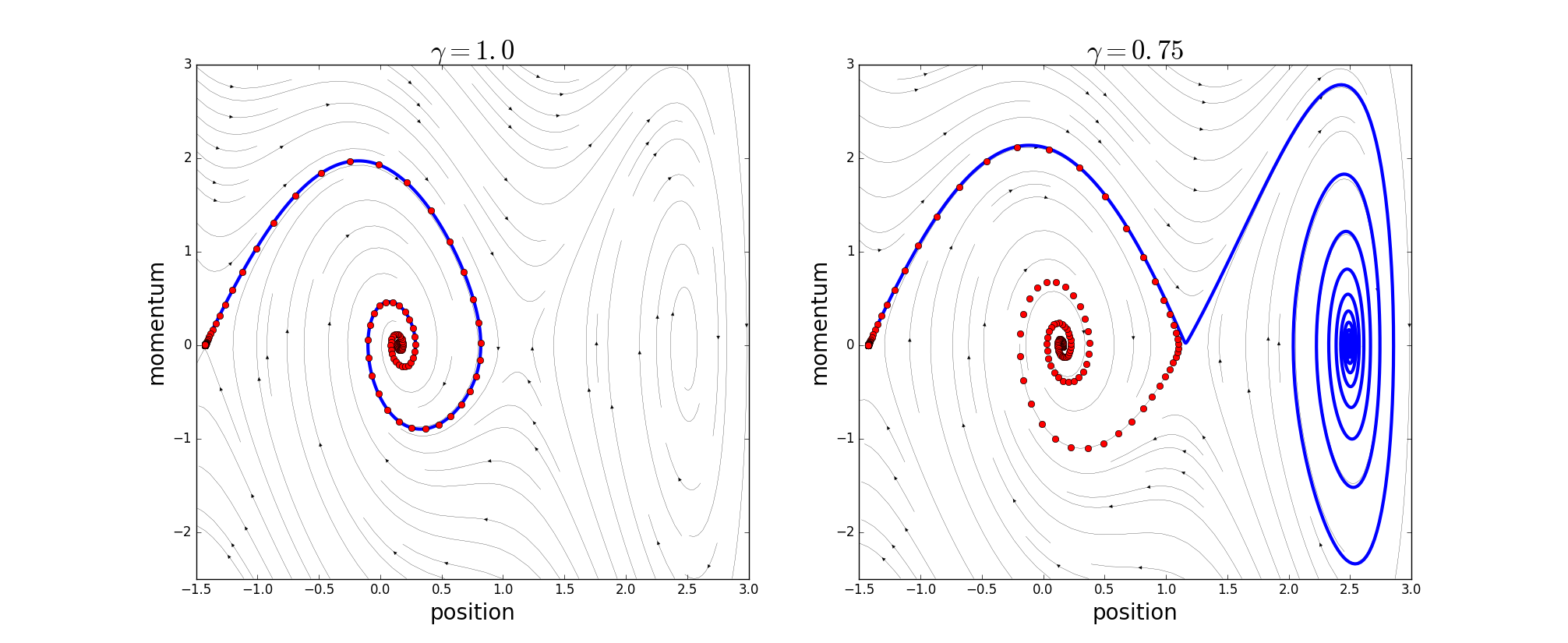}
\caption{[Color online] Solution to the boundary value problem (red dots) and the dynamical trajectory (blue) for the Langevin system with a sixth order polynomial potential. The black lines are the stream lines corresponding to the phase portrait of the system. On the left we have a friction coefficient of $\gamma = 1$, and the dynamical trajectory corresponds to minimizers of the Freidlin-Wentzell action. On the right the friction coefficient is $\gamma = 0.75$ and the dynamical trajectory no longer coincides with the action minimizer.}
\label{fig:phase}
\end{center}
\end{figure}

The deterministic path, streamlines, and a minimizer of the Freidlin-Wentzell action (computed by the method in \sref{sec:eul_lag}) for $\gamma=1$ and $\gamma=0.75$ are shown in \fref{fig:phase}. One can see that the deterministic path corresponding to $\gamma = 1$ (going downhill from the point $x_{AB}$) settles to the point $x_{B}$, and the minimizer and the deterministic path coincide.

When $\gamma = 0.75$ the deterministic path ends up in the well corresponding to location $x_C$ whereas the local minimizer of the Freidlin-Wentzell action ends directly  at $x_B$. The action value for the minimizing path for the $x_{AB}$ to $x_{B}$ transition  is less than $\sim 10^{-2}$, meaning that the most likely transition from $x_A$ to $x_B$ does not go through $x_C$, since this would correspond to an addition of $ 21.8046503$ to the action. The minimizer was found by solving the Euler-Lagrange equations via gradient descent (see \sref{sec:eul_lag}). For this case all of our attempts to utilize the string method failed to produce the correct global minimizer. All sufficiently well resolved strings that we tried had at least one of its points in the basin of attraction of the well corresponding to $x_C$. This seemed to be enough to sweep the rest of the string into a path that approximates the $x_{AB} \rightarrow x_{C} \rightarrow x_{BC} \rightarrow x_{B}$ transition, a suboptimal local minimizer of the Friedlin-Wentzell action.

 The third possibility, however, has not been theoretically ruled out. Nevertheless, our numerical computations of the $x_A \rightarrow x_B$ transition were insufficient to show significant advantage of crossing the separatrix away from the saddle.

\paragraph{Recommendation} The first recommendation is to compute the overdamped string and then simulate deterministic dynamics \sref{sec:saddle_points}. If the trajectories of the noiseless Langevin system end up in the same potential minima as was computed in the overdamped limit, then a good local minimizer of the action is found. If the trajectories end up in a potential minimum different from $x_A$ and $x_B$, then plan B is to apply the adapted string method \sref{sec:string_method}. If that again results in an insertion of an unwanted minimum in the middle of the transition (e.g., $x_A \rightarrow x_{AB} \rightarrow x_{C} \rightarrow x_{BC} \rightarrow x_{B}$ discussed above), the string method may have skipped a better action minimizer too. The last recourse then is to attempt to directly minimize the Freidlin-Wentzell action (see \sref{sec:eul_lag}). This, of course, presents additional difficulties since it may be computationally expensive (depending on the structure of the problem).

\subsection{Position dependent, matrix-valued friction coefficient}
\label{sec:position_dependent}
As the different effects of large and small scalar friction coefficients were described above, it is not difficult to see if the friction coefficient  is a constant matrix with both small and (relatively) large eigenvalues, then a mixed effect can be induced by this anisotropy, which further complicates the transition. Details of the case of a constant but anisotropic friction coefficient will no longer be individually discussed for conciseness; instead, in this section we show that the anisotropy only needs to be a local effect in order to result in a global change in minimizing trajectories. 

More precisely, consider an example with the potential
\begin{align}
U(x,y) &= \frac{1}{4}\left( x^2 - 1 \right)^2 - \cos( \omega  y ) / \omega^2,
\end{align}
where $\omega  = 10$.
The friction coefficient matrix is 
\begin{align}
\Gamma(x,y) &= \gamma
\begin{bmatrix}
1 & 0 \\
0 & 1
\end{bmatrix}
+ 
\gamma
e^{ -5\left( x^2 + y^2 \right) }
\begin{bmatrix}
c-1& c-\delta \\
c-\delta & c-1
\end{bmatrix}
\end{align}
where $c = 10$, $\delta  = 0.1$, and $\gamma = 0.1$. The anistropy is local with respect to the friction coefficient, but notably occurs at a saddle point of the system --- there the matrix is 
\begin{align}
\Gamma(0,0) &= 
\gamma
\begin{bmatrix}
c & c-\delta\\
c-\delta & c
\end{bmatrix}
\end{align}
whose eigenpairs are $\lambda =\gamma( 2c - \delta), v_{\lambda} = [1\text{ }  1 ]^T$ and $\lambda = \gamma \delta, v_{\lambda} = [1\text{ }  -1 ]^T$. There is a large disparity between the two eigenvalues.

\begin{figure}
\begin{center}
\includegraphics[width=1.0\textwidth]{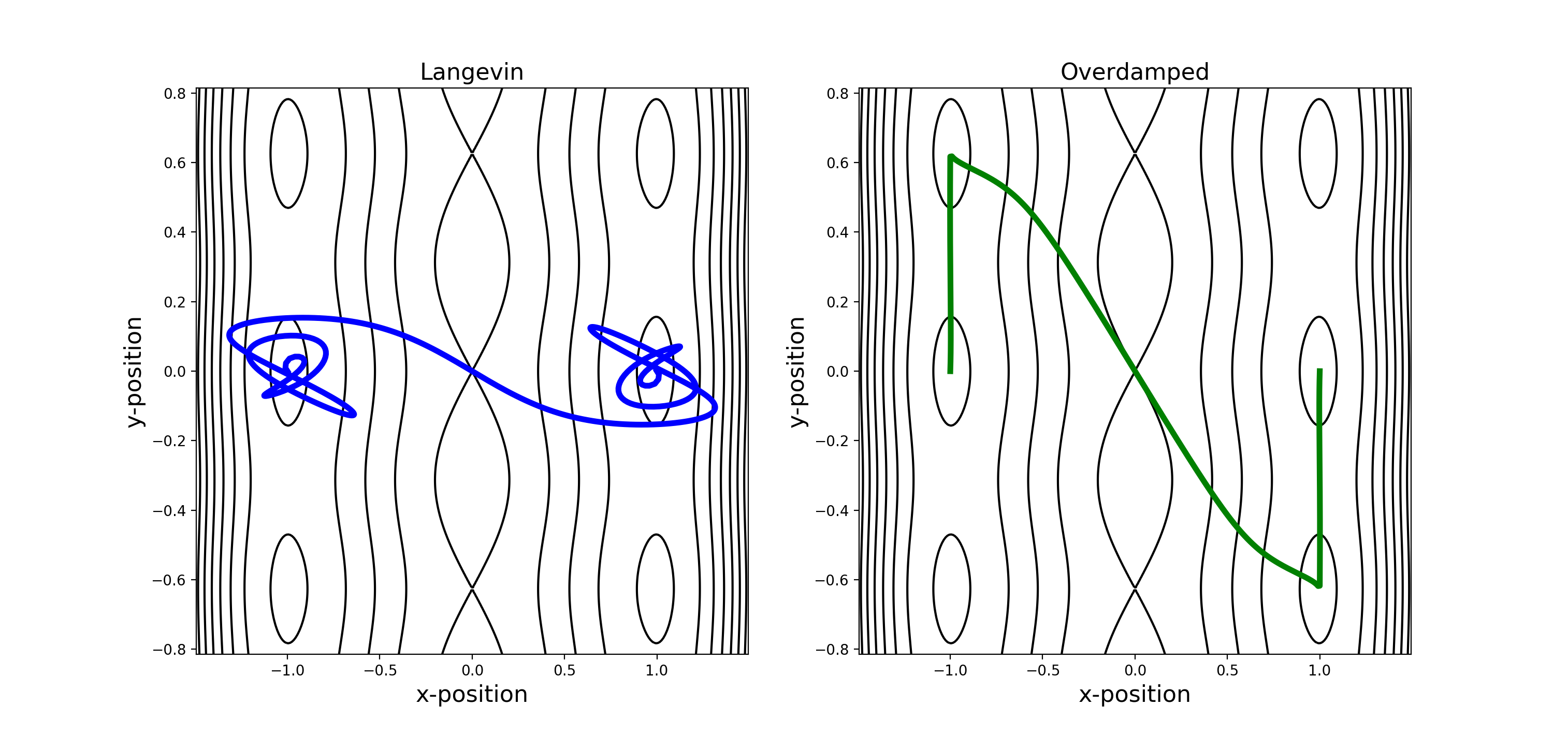}
\caption{[Color online] The Langevin and overdamped instantons. Even though there is only a local change in the friction coefficient, the change between instantons is global.}
\label{fig:position_dependent_tao}
\end{center}
\end{figure}

We consider the transition from $x_A = (-1,0)$ to $x_B = (1,0)$ (again with zero velocities). If the friction coefficient is isotropic (i.e., a constant scalar), the minimizing trajectories of both the overdamped and the Langevin system stay on the line connecting $x_A$ and $x_B$. The anisotropic position-dependent case is different. Utilizing the string method both adapted for the Langevin equation and for the overdamped equation leads to the scenarios in \fref{fig:position_dependent_tao}. The computed local action minimizer for the overdamped string ends up going through additional potential minimum (e.g., $x=-1, y=2\pi/\omega$) and saddle points. On the other hand the Langevin string still goes directly from $x_A$ to $x_B$ without passing through additional minima.

\begin{figure}
\begin{center}
\includegraphics[width=1.0\textwidth]{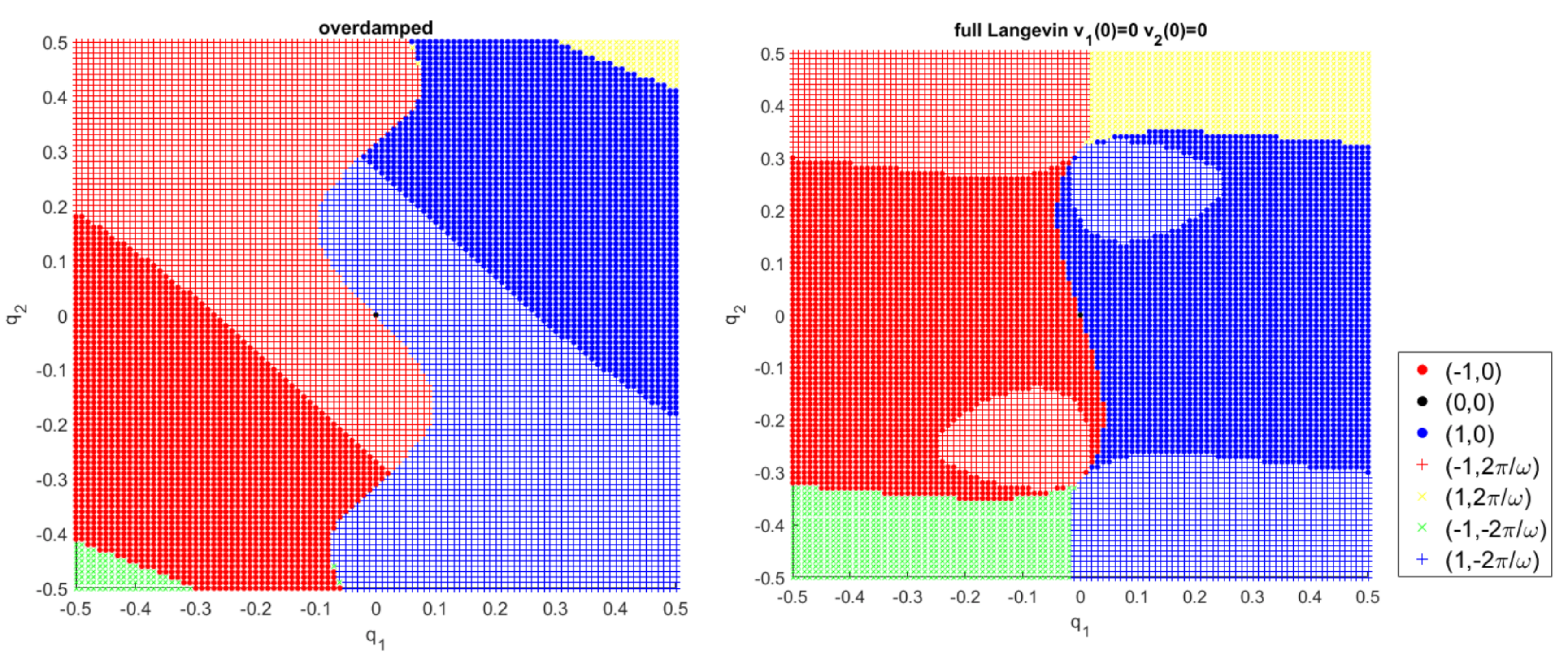}
\caption{[Color online] Numerical approximation of the basins of attraction for the position dependent friction coefficient  in the Langevin (left) and overdamped (right) systems, with an initial velocity $\mathbf{v} =0$ for the Langevin system. We also considered various other velocity initial conditions (not shown; available upon request) and were led to a similar structure as shown above.}
\label{fig:basin_position}
\end{center}
\end{figure}

Given that we cannot guarantee that the local minimizer for the overdamped functional is the global one, it is possible that a transition path that is more ``direct" exists for this system, but the simple heteroclinic dynamics \sref{sec:saddle_points} do not exist. An approximation to the basin of attraction for the Langevin and overdamped system is shown in \fref{fig:basin_position}. There we see that there is no heteroclinic connection from $(0,0)$ to $(1,0)$ in the noiseless overdamped system whereas it still exists Langevin system. At least two points are clear: (i) The straight line transition path, present when friction is isotropic, is no longer optimal due to the new friction, and (ii) overdamped transition and Langevin transition projected to position space may not coincide when friction coefficient is not uniformly large everywhere.

\paragraph{Recommendation} To numerically compute the instanton, here we give the same recommendation as in \sref{sec:underdamped}; however, caution should be used
when one tries to compute a transition involving more than two minima by the string method: a large number of points might be needed to represent the string to avoid degraded accuracy due to under-resolution; one fix is to compute pairwise transitions between minima, and then concatenate. In addition, if one utilizes the Euler-Lagrange equation, \ref{eq_ELspecific} needs to be modified because it was for the scalar constant coefficient case, but now we have position-dependence.

\paragraph{Two overdamped limits} It is also important to clarify, in the position-dependent friction case, what is an overdamped limit of 
\begin{equation}
\begin{array}{r@{}l}
 d q &= p dt  \\
 d p &= -\nabla U(q) dt - \Gamma(q) p dt+  \epsilon  \Gamma(q)^{1/2} d W .
 \end{array}
 \label{eq_positionDependentLangevinGeneral}
\end{equation}
We utilized the formal limit,
\begin{equation}
	dq=-\Gamma(q)^{-1} \nabla U(q) dt+\epsilon \Gamma(q)^{-1/2} dW.
	\label{eq_overdampedLimitRareEvent}
\end{equation}
Repeating the derivation in \sref{prob_form_sec}, it is not difficult to see that, under the condition of existences of needed heteroclinic orbits, the formal limit \ref{eq_overdampedLimitRareEvent} provides minimum action values consistent with those of the full system \ref{eq_positionDependentLangevinGeneral}. Because of this, we used this formal limit to investigate the rare event of metastable transition.

Note, however, that this formal limit does not conserve the $q$ marginal of the Boltzmann-Gibbs distribution, $Z^{-1}\exp(-(p^2/2+U(q))/(\epsilon^2/2)) \,dq\,dp$, which however remains as an invariant distribution of the Langevin equation \ref{eq_positionDependentLangevinGeneral} despite the no-longer-constant friction coefficient. One can show, by solving the stationary Fokker-Planck equation, that the following corrected overdamped limit preserves the marginal Boltzmann-Gibbs,
\begin{equation}
	dq=\left(-\Gamma(q)^{-1} \nabla U(q) + \frac{\epsilon^2}{2}\nabla A(q) \right) dt+\epsilon \Gamma(q)^{-1/2} dW,
	\label{eq_overdampedLimitRareEvent_Two}
\end{equation}
where $A(q):=\Gamma(q)^{-1}$. This equation is in fact consistent with the variable friction small mass limit in the literature, and we refer to \cite{small_mass_state_dependent} for more discussions (including convergence of the dynamics, not just long term statistics) and references.

We chose not to further investigate rare events in this correction \eqref{eq_overdampedLimitRareEvent} though, because its large deviation structure is in fact unclear (one cannot simply remove the $\nabla A$ term via $\epsilon\rightarrow 0$, since the noise also scales). An investigation could be interesting but it is out of the scope of this article. After all, our approach directly provided a solution to the full problem \ref{eq_positionDependentLangevinGeneral}.

\subsection{Velocity dependent, matrix-valued friction coefficient}
\label{sec:momentum_dependent}
\label{complex_sec}
Although complications previously investigated in \ref{sec:underdamped} and \ref{sec:position_dependent} can all add up to this case, to avoid redundancy we concentrate on the differences that a velocity dependence can induce. We do so by revisiting the Mueller potential with an anisotropic velocity dependent friction coefficient matrix in a mildly underdamped situation. When the friction coefficient depends on velocity, generally there is no analogous overdamped limit any more, and the system
\begin{equation}
\begin{array}{r@{}l}
 d q &= p dt  \\
 d p &= -\nabla U(q) dt - \Gamma(q,p) p dt+  \epsilon  \Gamma(q,p)^{1/2} d W .
 \end{array}
\end{equation}
may not admit Boltzmann-Gibbs as an invariant distribution. In terms of nonequilibrium statistics, we show the velocity dependence modifies instantons.

More specifically, consider an example where $\Gamma$ only depends on $p=[\dot{x}, \dot{y}]$, in the form of
\begin{align}
\label{velocity_dependent_matrix}
\Gamma( \dot{x} , \dot{y} ) &= 
\begin{bmatrix}
5 + \exp ( \dot{x} ) + \exp( \dot{y} ) & 1.25 \\
1.25  & 5 + \exp (-\dot{x} ) + \exp(-\dot{y})
\end{bmatrix}
\end{align}
Note $\Gamma( \dot{x} , \dot{y} )  \neq \Gamma( -\dot{x} ,- \dot{y} ) $; thus we need to pay special attention to the time-reversed Langevin equation as well, see \sref{prob_form_sec}. 

\begin{figure}
\begin{center}
\includegraphics[width=1.0\textwidth]{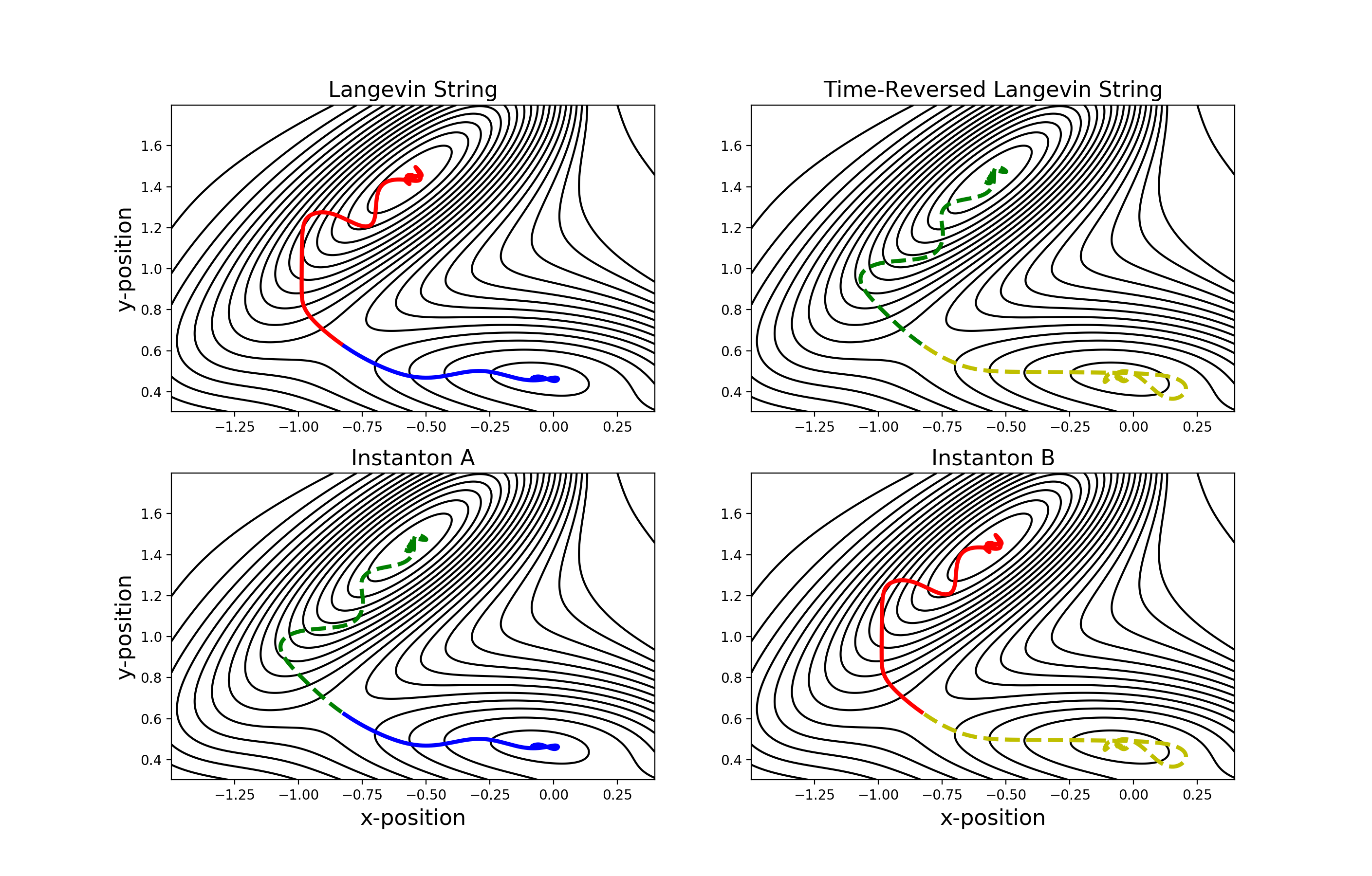}
\caption{[Color online] Strings and Instantons of the Langevin equation corresponding to a velocity dependent friction coefficient. For this case the strings are no longer instantons. The true instanton is a concatenation of the solid strings with the dashed strings and is different depending on whether or not one is interested in the transition from the top left well to the bottom right well (bottom left) or visa versa (bottom right).}
\label{fig:momentum}
\end{center}
\end{figure}

In \fref{fig:momentum} we compute the same transition from \sref{sec:overdamped}, but with \ref{velocity_dependent_matrix}. Here there is a difference between time-reversed dynamics and regular Langevin dynamics since the friction coefficient matrix is not symmetric with respect to sign reversal of velocity. The true minimizing path, is a concatenation of the solid curves and the dashed curves. The string method or dynamics can still be utilized to compute the minimizing trajectory, but twice the computation is necessary.

The difference between time-reversed dynamics and regular dynamics can lead to significant complications. Given the phenomena in \sref{sec:underdamped}, heteroclinic connections that exist in
\begin{align}
\ddot{\mathbf{x}} + \Gamma(\mathbf{\dot{x}}) \dot{ \mathbf{x} } + \nabla U  &= 0 
\end{align}
do not necessarily have an analog in
\begin{align}
\ddot{\mathbf{x}} + \Gamma(-\mathbf{\dot{x}}) \dot{ \mathbf{x} } + \nabla U  &= 0 .
\end{align}

\paragraph{Recommendation} Even though there is no analogous overdamped limit, one can still utilize knowledge of the string in the overdamped limit to compute instantons for this system. To make use of the overdamped limit, use the friction coefficient matrix $\Gamma(\mathbf{x},\mathbf{\dot{x}} = 0)$ and compute the overdamped string. This is, in a general setup, a position-dependent-friction string calculation as in \sref{sec:position_dependent}. Then one can use the method of dynamics \sref{sec:saddle_points} to compute trajectories for the Langevin equation and time-reversed Langevin equation. If the resulting noiseless (modified) Langevin trajectories do not end up at the target minimum, then one can also evolve two strings, one corresponding to Langevin dynamics and the other to the time-reversed dynamics, to compute a local action minimizer or obtain a better perturbation direction off the saddle. (Note that the method of obtaining an improved perturbation direction from the appendix is not always applicable to non-isotropic and non-constant friction coefficients.)  If the previous methods fail and if it is feasible to do so, one can try to compute minimizers utilizing the Euler-Lagrange equations.
 
 \section{Conclusion}
We analyzed minimizers of the Freidlin-Wentzell action for the inertial Langevin equation with respect to various types of friction coefficients, for illustrating features in inertial Langevin metastable transitions that differ from the familiar overdamped picture. We calculated local minimizers of the action functional utilizing three approaches, (i) an adaptation of string method, (ii) dynamics, and (iii) gradient descent minimization of the action, which not only depends on $\bold{x},\bold{\dot{x}}$ but also $\bold{\ddot{x}}$. We exploited the structure of the functional, which allows for a simple modification of the string method as well as an dynamics-based approach for improved resolution, provided heteroclinic connections exist. This structure also allows the calculation of instantons when the friction coefficient is position and/or velocity-dependent. When heteroclinic connections that exist for the overdamped system no longer correspond to that of the inertial system, we minimized the action directly, from whence we saw that instantons no longer need to obey dynamics or time-reversed dynamics. This implied that the minimum action values of the Langevin and overdamped functionals are different, leading to different transition rates.

\section{Acknowledgment}
MT is partially supported by NSF grant DMS-1521667 and ECCS-1829821 and AS is partially supported by DMS-1344199. MT is grateful for valuable discussions with Tony Leli\`evre and encouragements from Jianfeng Lu to continue exploring this interesting problem. We also appreciate comments from two anonymous reviewers, which greatly improved the manuscript.

 \appendix
 
 \section{Perturbation Direction}
 \label{perturb_direc}
 Suppose that $\mathbf{v}$ is an eigenvector of the matrices $A,B,C,D$ with eigenvalues $\lambda_A, \lambda_B, \lambda_C , \lambda_D$ respectively. Then
\begin{align}
\begin{bmatrix}
\alpha \mathbf{v} \\
\beta \mathbf{v}
\end{bmatrix}
\end{align}
is an eigenvector of the matrix
\begin{align}
\begin{bmatrix}
A & B \\
C & D
\end{bmatrix}
\end{align}
with eigenvalues
\begin{align}
\lambda  = 
\frac{\lambda_A + \lambda_D \pm \sqrt{(\lambda_A - \lambda_D)^2 + 4 \lambda_B \lambda_C}}{2} 
\end{align}
and 
\begin{align}
\frac{\beta}{\alpha } &= \frac{\lambda_A - \lambda_D \pm \sqrt{(\lambda_A - \lambda_D)^2 + 4 \lambda_B \lambda_C}}{2 \lambda_B} 
\end{align}
with appropriate modifications for the $\lambda_B = 0$ case.  The derivation is as follows: Starting with the ansatz for the eigenvector we compute
\begin{align}
\begin{bmatrix}
A & B \\
C & D
\end{bmatrix}
\begin{bmatrix}
\alpha \mathbf{v} \\
\beta \mathbf{v}
\end{bmatrix}
= 
\begin{bmatrix}
( \lambda_A + \frac{\beta}{\alpha}\lambda_B)\mathbf{I} & 0 \\
0 & ( \lambda_C\frac{\alpha}{\beta}+ \lambda_D ) \mathbf{I} 
\end{bmatrix}
\begin{bmatrix}
\alpha \mathbf{v} \\
\beta \mathbf{v}
\end{bmatrix}
\end{align}
and we choose the parameters $\alpha$ and $\beta$ such that the diagonals of the matrix are equal.

With respect to the Langevin equation with a constant friction coefficient matrix $\Gamma$, and Hessian of the potential $H$, the Hessian of the Langevin system is the following block matrix
\begin{align}
H_L  = \begin{bmatrix}
0 & \mathbf{I} \\
-H & -\Gamma 
\end{bmatrix}.
\end{align}
Assume that $\Gamma$ and the $H$ share an eigenvector  $\mathbf{v}$, as for example would be the case for an isotropic friction coefficient. Based off of the previous calculations the eigenvector and values are 
\begin{align}
\mathbf{ v}_L &= \begin{bmatrix}
 \mathbf{v} \\
\beta  \mathbf{v}
\end{bmatrix}
\end{align}
where \begin{align}
\alpha &= 1 \\
\beta &= \frac{-\lambda_\Gamma + \sqrt{(\lambda_\Gamma )^2 - 4 \lambda_H} }{2} \\
\end{align}
and $\beta$ is an eigenvalue. Along an unstable direction $\lambda_H < 0$ and in the limit $\lambda_\Gamma \rightarrow \infty$ we have 
$\beta = \frac{\sqrt{-\lambda_H}}{\lambda_\Gamma} + \mathcal{O}\left( \frac{\lambda_H}{(\lambda_\Gamma)^2}\right)$.

\section*{References}

\bibliography{./mybibfile_new}

\end{document}